\def\apj{ApJ}
\def\apjl{ApJL}
\def\mnras{MNRAS}
\def\aap{A\&A}

\def\simlt{\lower.5ex\hbox{$\; \buildrel < \over \sim \;$}}
\def\simgt{\lower.5ex\hbox{$\; \buildrel > \over \sim \;$}}
\def\simpt{\lower.5ex\hbox{$\; \buildrel \propto \over \sim \;$}}
\def\mag{\mbox{ mag}}
\def\kms{\mbox{ km s$^{-1}$}}
\def\mpc{\mbox{ Mpc}}
\def\kpc{\mbox{ kpc}}
\def\pc{\mbox{ pc}}

\def\msun{\mbox{ M}_\odot}

\documentclass{iaus}
\usepackage{graphicx}

\title[Substructure and Cosmology]{Small-Scale Structure, Missing Galaxies              and Gravitational Lensing}

\author{R. Benton Metcalf\footnote{Hubble Fellow}}

\affiliation{Department of Astronomy \& Astrophysics, University of California,
Santa Cruz, CA 95064, USA email: bmetcalf@ucolick.org}

\pubyear{2004}
\volume{225}
\pagerange{1--8}
\date{?? and in revised form ??}
\jname{Impact of Gravitational Lensing on Cosmology}
\editors{Mellier, Y. \& Meylan,G. eds.}
\begin{document}

\maketitle

\begin{abstract}
The gravitational lensing constraints on the small mass end of the
$\Lambda$CDM mass function are discussed.
Here a conservative approach is taken where only the most difficult to explain
image flux anomalies in strong QSO lenses are emphasized.
Numerical simulations are performed to compare predictions for the
$\Lambda$CDM small scale mass function with the observed flux ratios.
It is found that the cusp caustic lens anomalies and the
disagreements between monochromatic flux ratios and simple lens models can be 
explained without any substructure in the primary lenses' dark matter
halos.  Extragalactic $\Lambda$CDM halos of mass $< 10^{9}\msun$ are
enough to naturally explain these anomalies.  This does not mean that
substructure within the host lens is not contributing.  In fact, it
could dominate the lensing depending on how much of substructure survives in the
centers of galactic halos. 

Spectroscopic gravitational lensing provides more information on the
nature of these substructures.  Thus far, the one relevant case in which this
technique has been used, observations of Q2237+0305, shows evidence of
more small mass halos ($\sim 10^6\msun$) than is expected in the $\Lambda$CDM model.
\end{abstract}

\firstsection 
\section{Introduction}
\label{sec:introduction}
The Cold Dark Matter (CDM) model predicts a large quantity of small mass dark
matter halos ($\simlt 10^7 \msun$) that must have little or no stars in
them to agree with the number counts of dwarf galaxies.  Quasars
(QSOs) that are being gravitationally lensed into multiple images have
recently been used to put limits on the surface density and mass of such invisible
subclumps 
(\cite{1998MNRAS.295..587M}; \cite[Metcalf \& Madau 2001]{2001ApJ...563....9M};
\cite{2002ApJ...565...17C}; \cite[Metcalf 2002]{2002ApJ...580..696M};
\cite{2002ApJ...567L...5M}; \cite{Dalal2002}; \cite{astro-ph/0112038};
\cite{2003ApJ...584..664K}; \cite{cirpass2237}).
The question arises as to whether these observations are
reliable and compatible with the current $\Lambda$CDM model.

Some lenses provide much stronger and more certain constraints on the
small scale structure than others.  I try to take a conservative
approach here and consider only the lenses that provide clean,
relatively unambiguous constraints. 
The standard $\Lambda$CDM cosmological model will have the
cosmological parameters $\Omega_m=0.3$, $\Omega_\Lambda=0.7$,
$\sigma_8=0.9$, $H_o=70 \kms\mpc^{-1}$ and a scale free initial power spectrum.

\section{Some Background}

There are essentially four ways that have been proposed for detecting
substructure in multiply imaged QSO lenses.  They are briefly
described here.

\subsection{Monochromatic magnification ratio anomalies}
\label{sec:monochr-magn-ratio}

It was proposed by \cite{2001ApJ...563....9M} that the missing CDM
substructure could be searched for by comparing the flux ratios of
4-image QSO lenses with those predicted by lens models.
Simulations showed that if the substructure has a small mass scale
the image positions can be used to constrain the host, smooth lens
model.  It was subsequently shown that the magnification ratios of
observed lenses generically do not agree with simple lens models
(\cite{2002ApJ...567L...5M,2002ApJ...565...17C,Dalal2002}).  These
anomalies are probably the result of substructure (see \cite[Kochanek
  \& Dalal (2003)]{KD2003}
for some arguments), but when
interpreting the results degeneracies in the lens models become a
problem.  More complicated lens models can fit the image positions
just as well and give different predictions for the magnification
ratios.  To actually measure properties of these substructures a more
precise method is required.

\subsection{The cusp caustic relation}
\label{sec:cusp-caust-relat}
It can be shown that if the source is close to a cusp
in the caustic of a sufficiently smooth lenses three of the images
will be clustered together and the magnifications of the close triplet
will sum to zero; taking the parity reversed images to have negative
magnification (\cite{1992A&A...260....1S}).  To make this prediction
independent of the intrinsic luminosity of the  
QSO the images in the triplet are labeled A through C and the cusp
caustic parameter, $R_{\rm cusp}$, is defined as
\begin{equation}
R_{\rm cusp}\equiv \frac{\mu_A+\mu_B+\mu_C}{|\mu_A|+|\mu_B|+|\mu_C|}
\end{equation}
which should be zero if the lens map is sufficiently smooth.
Small scale structure on approximately the scale 
of the image separations will cause $R_{\rm cusp}$ to differ from zero
fairly independently of the form of the rest of the lens 
(\cite{1998MNRAS.295..587M,KGP2002}).

The five well observed cusp caustic lenses all show violations of
the magnification relation at some level.  Two of these are only
measured in the optical/near-IR where microlensing by stars could be important.
The three with $R_{\rm cusp}$ measured in the radio (B0712+472,
B2045+265 and B1422+231) clearly violate the relation although
B1422+231 is less clear than the others.  In section~\ref{sec:results}
(and in \cite{Metcalf04,AMC2004}) the significance of these
violations is investigated.

\subsection{Spectroscopic gravitational lensing}
\label{sec:spectr-grav-lens}

It was proposed by \cite{MM02} that much of the lens model degeneracy
can be removed and the sensitivity to substructure properties
improved by utilizing the fact that the different emission regions of the
source QSO have different physical sizes.  If the lens is smooth on
the scales that bridge the sizes of the emission regions, the
magnification of those regions should be the same and thus the
magnification ratios should be the same.  The visible and
near-infrared (near-IR)
continuum emission regions are small ($\sim$ 100~AU) and their
magnification can be affected by microlensing by ordinary stars in the
lens galaxy.  The broad line emission region is $\sim 0.1\pc$ in size
and is less affected by microlensing in most cases.  The radio and
mid-IR regions are $\sim 10\pc$; their magnification will be
dominated by larger scales than stars.  The narrow line emission region
is even larger, $\simgt 100\pc$.  The magnification ratios in these bands and lines
can be compared to constrain the mass, concentration and number
density of substructures.  \cite{cirpass2237} found that the
narrow line magnification ratios do not agree with the radio and
mid-IR ratios (although the radio and mid-IR ratios do agree with each
other) in the lens Q2237+0305.
It was shown in that paper that at least a few \%
of the surface density of the lens needs to be in substructure of
mass $\simlt 10^7\msun$ to explain this mismatch.  This result is not
consistent with the present $\Lambda$CDM predictions in that it
requires too much mass in 
very small mass halos either inside the lens of somewhere along
the line of sight.  However, these predictions could be significantly
underestimating the amount of substructure because of numerical
effects in the simulations (see J. Taylor, these proceedings).

\subsection{Bent radio jets}

It is also possible to look for substructure by comparing the
curvature, on milliarcsecond scales, of multiply imaged radio jets
(\cite[Metcalf \& Madau 2001]{2001ApJ...563....9M}).  There is some evidence that the bends
in the jets of B1152+199 are not compatible with a smooth lens, but the
case is not yet water tight (\cite[Metcalf 2002]{2002ApJ...580..696M}).

\section{Simulations}
\label{sec:simulations}

Numerical simulations are necessary to calculate the expected
influence of small scale structure on the magnification ratios.     
Generally there are multiple small halos affecting a single
image, the size of the source (in the radio, mid-IR or narrow
lines) is significant compared to the sizes of the substructures and
the effect of a single substructure on multiple images must be considered.
Any massive object near the line of sight inside or outside of the primary lens
could potentially contribute.  Both
contributions have been simulated, but here we concentrated on the
extragalactic part.  The simulation method is more thoroughly discussed in
\cite{Metcalf04}.  Here it is briefly outlined.

The large number of small halos and the large range in size scales,
from the size of the primary lens ($\sim 100\kpc$) to the size of the source
($\simlt 0.1\pc$ for the broad line emission region), make finding the images
and calculating magnifications challenging and time consuming.  An
adaptive mesh refinement technique is used to overcome these problems.  
The entire lens is simulated at once in all cases.

For extragalactic halos the Press-Schechter (PS) formalism is used to
calculate the mass function from which a random sample of halos is drawn.
In this mass regime the PS mass function differs very little from
the Sheth-Tormen mass function.   The structure of these halos is
taken to be of the NFW form truncated at the virial radius.  The
initial power spectrum is taken to be scale invariant and normalized
to $\sigma_8=0.9$.  The concentrations of the halos are set according
to 
\begin{equation}
c=c_o\left(\frac{M}{10^{12}\msun}\right)^{-\beta}
\end{equation}
with $c_o\simeq 12$ and $\beta =0.13$, in agreement with
\cite[Zentner \& Bullock (2003)]{astro-ph/0304292}.
In addition to the substructure, a model for the host lens must be
chosen.    A SIE + external shear model is used in these simulations.
The lensing results can change significantly when $\beta$ or
$\sigma_8$ is changed.

\section{Results of simulations}
\label{sec:results}

\begin{figure}[t]
\begin{center}
 \includegraphics[height=3.in]{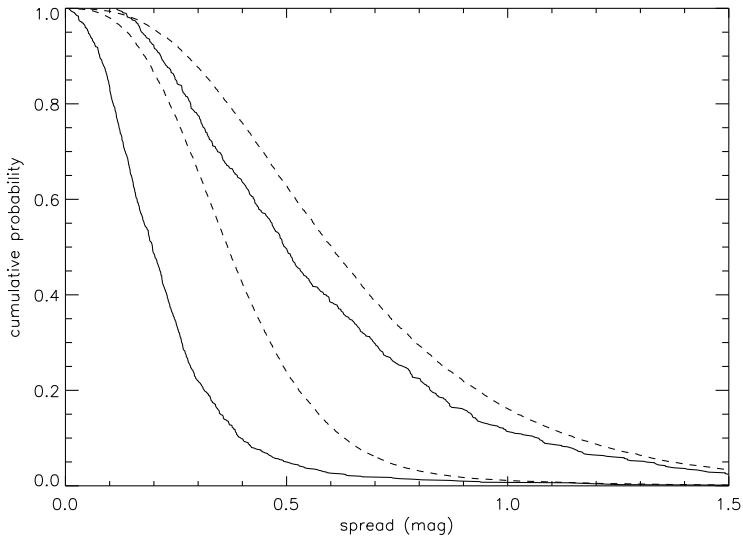}
\caption[the]{\footnotesize This is the probability of having
a magnification ratio disagree with the lens model by more than a
certain magnitude for Q2237+0305.  The two solid curves are without
observational noise and the dashed curves are with 0.15~mag of noise.
For each type of curve the one on the left is for extragalactic halos
with $10^7\msun < m < 10^8\msun$ and the one on the right is for
$10^7\msun < m < 10^9\msun$.  There is no substructure inside the
primary lens.}
\label{fig:spread_ratio}
\end{center}
\end{figure}

Simulations were performed to mimic the observed lenses with the addition of
$\Lambda$CDM halos.  The resulting combinations of image
magnifications are then compared with those observed to determine if
the observed anomalies are expected to be reasonably common in this
cosmological model or unlikely.

To represent lenses in the Einstein cross
configuration, a host lens model is constructed that fits
the image positions of Q2237+0305.  The effects of substructure within
the host lens and its contributions to {\it spectroscopic lensing} were
investigated in \cite{cirpass2237}.  Only the extragalactic
contribution to {\it monochromatic} magnification ratios is discussed here.

Figure~\ref{fig:spread_ratio} shows a cumulative distribution of the
largest discrepancy (in magnitudes) between the smooth model
predictions and simulated values for the three magnification ratios.
The source size is 1~pc in 
this case.  Most of the anomalies are caused by the high end of the mass
distribution, $m\simeq 10^8-10^9\msun$.  
One can see that these discrepancies are
rather large even without any substructure in the host lens itself --
discrepancies as large as $\sim 0.5\mag$ are expected in half the
cases.  The typical discrepancies between observed flux ratios and
models are a few tenths of a magnitude
(see \cite{2002ApJ...567L...5M,KD2003}).  This makes the observed
{\it monochromatic} ratio anomalies consistent with $\Lambda$CDM,
simple lens models and no substructure internal to the primary lenses.  

$\Lambda$CDM halos seem easily capable of changing the monochromatic
magnification ratios by this much, but they do not produce the
mismatch in the magnifications of different size sources as seen by
\cite{cirpass2237}.  This 
problem can be traced to a deficiency of small mass ($\sim 10^6$)
halos in the $\Lambda$CDM model.

\begin{figure}[t]
\begin{center}
\includegraphics[height=3.in]{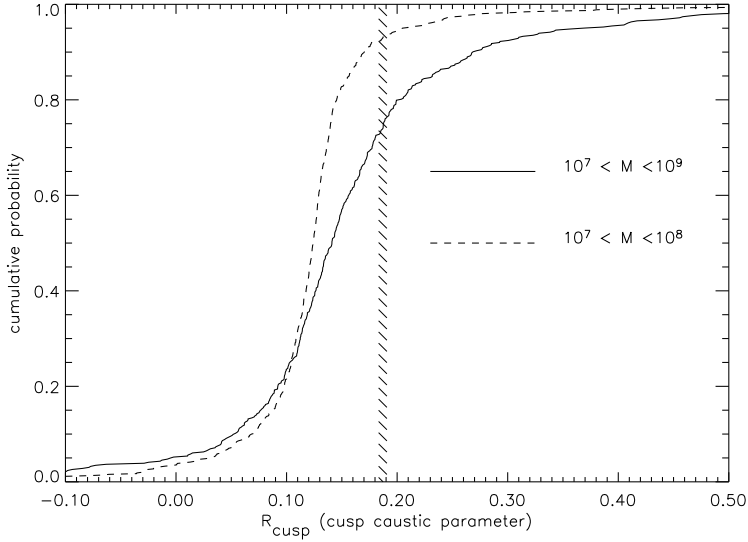}
\caption[the]{\footnotesize The distribution of the cusp caustic
  parameter, $R_{\rm cusp}$, for lens B1422+231 with only
  extragalactic standard $\Lambda$CDM small-scale structure.  The
  observed value in the radio with error is shown as the hashed
  region.  The different curves correspond to the halos mass ranges
  shown.  It can be seen that most of the changes in $R_{\rm cusp}$
  are caused by relatively large mass halos, $10^8\msun<m<10^9\msun$.
  There is about a 25\% chance of $R_{\rm cusp}$ differing from zero
  by more than is observed.}
\label{fig:r_prob1422}
\end{center}
\end{figure}

To investigate violations of the cusp caustic relation simulations
were done for several models designed to mimic observed lenses.
Figure~\ref{fig:r_prob1422} shows the
distribution of $R_{\rm cusp}$ for B1422+231 with the expected population of
extragalactic halos only.  The first thing to note is the marked
asymmetry in the distribution.  As previously seen
(\cite[Metcalf 2001]{astro-ph/0109347}; \cite[Matcalf \& Madau
  2001]{2001ApJ...563....9M}; \cite{2002ApJ...580..685S}), the
magnifications of negative magnification images are affected by
substructure differently than positive magnification images.
When substructure is added, $R_{\rm cusp}$ is biased toward positive
values. 

Also shown in figure~\ref{fig:r_prob1422} is the observed value of
$R_{\rm cusp}$ for comparison.  There is a perfectly reasonable
probability of $\simeq 0.28$ that $R_{\rm cusp}$ would be even
larger than the observed value.  By comparing the two different ranges
for the halo masses, it can be seen that violations in the cusp caustic
relation are mostly caused by more massive halos in this case.  In
light of this, the violation of 
the cusp caustic relation in B1422+472 seems fully consistent with
the $\Lambda$CDM model even without substructure within the halo of
the primary lens.

We can also compare figure~\ref{fig:r_prob1422} to lens B0712+472
which has a similar configuration to B1422+231 although a lower source
redshift.  It is easily seen that its value of $R_{\rm cusp}=0.26\pm
0.02$ is not particularly unlikely (there is a $\sim 12 \%$ probability of it
being larger) and thus does not require an additional
explanation beyond the expected population of extragalactic halos.

\begin{figure}[t]
\begin{center}
\includegraphics[height=3.0in]{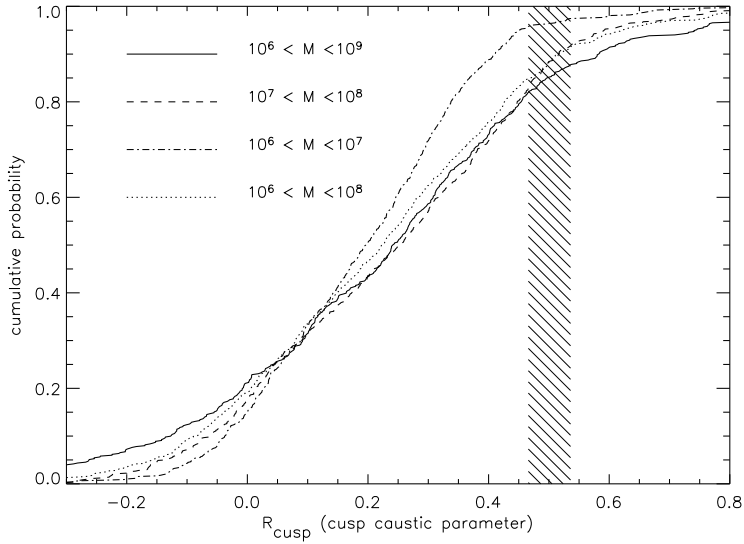}
\caption[the]{\footnotesize The cumulative distribution for $R_{\rm
    cusp}$ in the tight long axis case like B2045+265 with only
  extragalactic substructure.  The observed value of $R_{\rm cusp}$ in
the radio is shown by the hashed region.  The included halo mass
ranges are shown.}
\label{fig:r_prob2045cosmic}
\end{center}
\end{figure}

Lens B2045+265 is a more extreme cusp caustic case.  
Figure~\ref{fig:r_prob2045cosmic} shows the results for simulations
with just extragalactic $\Lambda$CDM halos.  With a halo mass range of
$10^6\msun<m<10^9\msun$ the observed $R_{\rm cusp}$ does not appear
strongly disfavored -- $15\%$ chance of it being larger.
Considering the additional substructure within the host lens, the
observed $R_{\rm cusp}$ seems perfectly consistent with $\Lambda$CDM.

\section{Conclusion}
\label{sec:discussion}

It has been shown here that anomalies in the monochromatic (as opposed
to differential) magnification ratios of
cusp caustic lenses can all be explained naturally within the
$\Lambda$CDM model with little if any substructure within the dark
matter halo of the primary lenses.  Extragalactic halos are enough to
account for these anomalies.  Furthermore, the typical observed
anomalies in the monochromatic magnification ratios of 
several tenths of a magnitude  -- when compared to simple lens models
-- are easily explained in the same way.  
The contribution to flux anomalies from extragalactic halos is
found to be significant.  Measuring the amount of substructure that is within
the primary lens halos for comparison with Nbody simulations will
require a large number of lenses and an 
accurate prediction for the extragalactic contribution.  The
extragalactic population of halos with masses $<10^8\msun$ has not
been directly observed either, even in the relatively local universe. 
The anomalies in the monochromatic magnification ratios could also
be explained by smaller scale structures than cansidered here since
they do not provide significant constraints on the substructure mass. 

It is significant that  all of
the observed cusp caustic parameters, $R_{\rm cusp}$, are positive.
In light of the marked asymmetry in the distributions of $R_{\rm cusp}$
from the simulations, the positive values can be seen as
further support for the conclusion that these anomalies are being caused by
some kind of small scale structure.

In contrast to the monochromatic magnification ratios, the
spectroscopic gravitational lensing observations of Q2237+0305
require more small mass halos than are expected in the $\Lambda$CDM model.
Bent multiply imaged radio jets also hint, although less securely, at
a large number of small mass objects.  The case for small mass
substructure is not yet secure, but further data should resolve the issue.
On the
theoretical side, advances in cosmological simulations should soon make it
possible to extend predictions for the mass function
of substructures within the halos of large galaxies down to smaller
masses and smaller galactocentric radii where they can be more
directly compared with observations.   At this time, there is an
inconsistency that needs to be resolved between the $\Lambda$CDM model
and the gravitational lensing observations. 

\begin{acknowledgments}
Financial support was provided by NASA through Hubble Fellowship 
grant HF-01154.01-A awarded by the Space Telescope Science Institute,
which is operated by the Association of Universities for Research 
in Astronomy, Inc., for NASA, under contract NAS 5-26555
\end{acknowledgments}

\end{document}